\documentstyle[12pt]{article}

\newcommand{\sect}[1]{\setcounter{equation}{0}\section{#1}}

\def\bseq{\begin{subequation}}  
\def\eseq{\end{subequation}}
\def\bsea{\begin{subeqnarray}}  
\def\esea{\end{subeqnarray}}

\newcommand{\beq}{\begin{equation}}
\newcommand{\eeq}{\end{equation}}
\newcommand{\bea}{\begin{eqnarray}}
\newcommand{\eea}{\end{eqnarray}}
\newcommand {\non}{\nonumber}

\renewcommand{\a}{\alpha}
\renewcommand{\b}{\beta}

\newcommand{\th}{\theta}

\newcommand{\pa}{\partial}
\newcommand{\di}{\partial}
\newcommand{\g}{\gamma}
\newcommand{\G}{\Gamma}

\renewcommand{\L}{\Lambda}

\newcommand{\f}{\phi}
\newcommand{\vf}{\varphi}
\newcommand{\F}{\Phi}

\newcommand{\vp}{\varpi}

\newcommand{\s}{\sigma}
\renewcommand{\S}{\Sigma}

\renewcommand{\o}{\omega}
\renewcommand{\O}{\Omega}

\newcommand{\chib}{\bar{\chi}}

\newcommand{\calX}{{\cal X}}

\def\Mb{\kern 2pt\mathchoice
            {
             \vbox{\hrule width10pt height 0.4pt depth 0pt
                 \kern 1.2pt\hbox{\kern -2pt$\displaystyle M$}}}
            {
                 \vbox{\hrule width10pt height 0.4pt depth 0pt
                 \kern 1.2pt\hbox{\kern -2pt$\textstyle M$}}}
            {
\vbox{\hrule width6pt height 0.4pt depth 0pt
                 \kern 1.0pt\hbox{\kern -2pt$\scriptstyle M$}}}
            {
                 \vbox{\hrule width5pt height 0.4pt depth 0pt
                 \kern 0.8pt\hbox{\kern -2pt$\scriptscriptstyle M$}}}}

\def\Sb{\kern 2pt\mathchoice
            {
                 \vbox{\hrule width6pt height 0.4pt depth 0pt
                 \kern 1.2pt\hbox{\kern -2pt$\displaystyle S$}}}
            {
                 \vbox{\hrule width6pt height 0.4pt depth 0pt
                 \kern 1.2pt\hbox{\kern -2pt$\textstyle S$}}}
            {
                 \vbox{\hrule width3.5pt height 0.4pt depth 0pt
                 \kern 1.0pt\hbox{\kern -2pt$\scriptstyle S$}}}
            {
                 \vbox{\hrule width3pt height 0.4pt depth 0pt
                 \kern 0.8pt\hbox{\kern -2pt$\scriptscriptstyle S$}}}}

\def\Rb{\kern 2pt\mathchoice
            {
                 \vbox{\hrule width5.5pt height 0.4pt depth 0pt
                 \kern 1.2pt\hbox{\kern -2.5pt$\displaystyle R$}}}
            {
                 \vbox{\hrule width5.5pt height 0.4pt depth 0pt
                 \kern 1.2pt\hbox{\kern -2.5pt$\textstyle R$}}}
            {
                 \vbox{\hrule width3.5pt height 0.4pt depth 0pt
                 \kern 1.0pt\hbox{\kern -2.2pt$\scriptstyle R$}}}
            {
                 \vbox{\hrule width3pt height 0.4pt depth 0pt
                 \kern 0.8pt\hbox{\kern -2.2pt$\scriptscriptstyle R$}}}}

  \def\pp{{\mathchoice
          {
              \kern 1pt%
              \raise 1pt
              \vbox{\hrule width5pt height0.4pt depth0pt
                    \kern -2pt
                    \hbox{\kern 2.3pt
                          \vrule width0.4pt height6pt depth0pt
                          }
                    \kern -2pt
                    \hrule width5pt height0.4pt depth0pt}%
                    \kern 1pt
           }
            {
              \kern 1pt%
              \raise 1pt
              \vbox{\hrule width4.3pt height0.4pt depth0pt
                    \kern -1.8pt
                    \hbox{\kern 1.95pt
                          \vrule width0.4pt height5.4pt depth0pt
                          }
                    \kern -1.8pt
                    \hrule width4.3pt height0.4pt depth0pt}%
                    \kern 1pt
            }
            {
              \kern 0.5pt%
              \raise 1pt
              \vbox{\hrule width4.0pt height0.3pt depth0pt
                    \kern -1.9pt  
                    \hbox{\kern 1.85pt
                          \vrule width0.3pt height5.7pt depth0pt
                          }
                    \kern -1.9pt
                    \hrule width4.0pt height0.3pt depth0pt}%
                    \kern 0.5pt
            }
            {
              \kern 0.5pt%
              \raise 1pt
              \vbox{\hrule width3.6pt height0.3pt depth0pt
                    \kern -1.5pt
                    \hbox{\kern 1.65pt
                          \vrule width0.3pt height4.5pt depth0pt
                          }
                    \kern -1.5pt
                    \hrule width3.6pt height0.3pt depth0pt}%
                    \kern 0.5pt
            }
        }}

  \def\mm{{\mathchoice
                       {
                             \kern 1pt
               \raise 1pt    \vbox{\hrule width5pt height0.4pt depth0pt
                                  \kern 2pt
                                  \hrule width5pt height0.4pt depth0pt}
                             \kern 1pt}
                       {
                            \kern 1pt
               \raise 1pt \vbox{\hrule width4.3pt height0.4pt depth0pt
                                  \kern 1.8pt
                                  \hrule width4.3pt height0.4pt depth0pt}
                             \kern 1pt}
                       {
                            \kern 0.5pt
               \raise 1pt
                            \vbox{\hrule width4.0pt height0.3pt depth0pt
                                  \kern 1.9pt
                                  \hrule width4.0pt height0.3pt depth0pt}
                            \kern 1pt}
                       {
                           \kern 0.5pt
             \raise 1pt  \vbox{\hrule width3.6pt height0.3pt depth0pt
                                  \kern 1.5pt
                                  \hrule width3.6pt height0.3pt depth0pt}
                           \kern 0.5pt}
                       }}

\def\pd{{\kern0.5pt
                   + \kern-5.05pt \raise5.8pt\hbox{$\textstyle.$}\kern
0.5pt}}

\def\pmd{{\kern0.5pt
                  \pm \kern-5.05pt \raise6.3pt\hbox{$\textstyle.$}\kern1.5pt}}

\def\md{{\mathchoice
   {
      {{\kern 1pt - \kern-6.2pt \raise5pt\hbox{$\textstyle.$}\kern 1pt}}}
    {
      {{\kern 1pt - \kern-6.2pt \raise5pt\hbox{$\textstyle.$}\kern 1pt}}}
    {
      {\kern0.5pt - \kern-5.05pt \raise3.4pt\hbox{$\textstyle.$}\kern0.5pt}}
    {
      {\kern0.5pt - \kern-5.05pt \raise3.4pt\hbox{$\textstyle.$}\kern0.5pt}}}}

\newcommand{\Eh}{\hat{E}}

\newcommand{\Apm}{A_+^{~-}}
\newcommand{\Amp}{A_-^{~+}}
\newcommand{\Apmd}{A_{\pd}^{~\md}}
\newcommand{\Ampd}{A_{\md}^{~\pd}}
\newcommand{\Apdmd}{A_{\pd}^{~\md}}
\newcommand{\Amdpd}{A_{\md}^{~\pd}}

\newcommand{\BAH}{\buildrel \leftarrow \over H}

\newcommand{\ad}{{\dot{\alpha}}}
\newcommand{\bd}{{\dot{\beta}}}
\newcommand{\Del}{\nabla}
\newcommand{\Delb}{\bar{\nabla}}
\newcommand{\Delp}{\nabla_{+}}
\newcommand{\Delm}{\nabla_{-}}
\newcommand{\Delpd}{\nabla_{\pd}}
\newcommand{\Delmd}{\nabla_{\md}}
\newcommand{\Delpp}{\nabla_{\pp}}
\newcommand{\Delmm}{\nabla_{\mm}}
\newcommand{\Delsq}{\nabla^2}
\newcommand{\Delbsq}{{\bar{\nabla}}^2}
\newcommand{\Dela}{\nabla_{\alpha}}
\newcommand{\Delad}{\nabla_{\dot{\alpha}}}

\newcommand{\bD}{{\bf D}}

\newcommand{\bDpp}{{\bf D}_{\pp}}
\newcommand{\bDmm}{{\bf D}_{\mm}}

\newcommand{\calD}{{\cal D}}

\newcommand{\calDpp}{{\cal D}_\pp}
\newcommand{\calDmm}{{\cal D}_\mm}

\newcommand{\reff}[1]{(\ref{#1})}

\newcommand{\sihalf}{{\frac{i}{2}}}
\newcommand{\half}{\frac{1}{2}}
\newcommand{\ihalf}{\frac{i}{2}}

\renewcommand{\thefootnote}{\fnsymbol{footnote}}

\begin{document}

\newpage
\begin{titlepage}
\begin{flushright}
{hep-th/9508139}\\
{BRX-TH-378}
\end{flushright}
\vspace{2cm}
\begin{center}
{\bf {\large  SUPERSPACE MEASURES,  INVARIANT ACTIONS, AND COMPONENT
PROJECTION FORMULAE  FOR (2,2)
SUPERGRAVITY}}\\
\vspace{1.5cm}
Marcus T. Grisaru\footnote{
Work partially supported by the National Science Foundation under
grant PHY-92-22318.} \\
and\\
\vspace{1mm}
Marcia E. Wehlau\footnote{\hbox to \hsize{Current address:
Mars Scientific Consulting, 28 Limeridge Dr.,
 Kingston, ON CANADA K7K~6M3}}\\
\vspace{1mm}
{\em Physics Department, Brandeis University, Waltham, MA 02254, USA}

\vspace{1.1cm}
{{ABSTRACT}}
\end{center}

\begin{quote}
In the framework of the prepotential description of superspace
two-dimensional $(2,2)$ supergravity,
we discuss the construction of invariant integrals. In addition to the full
superspace
measure, we derive the measure for chiral superspace, and obtain the
 explicit
expressions for going from superspace actions to component actions.
We consider both the minimal $U_A(1)$ and the extended
 $U_V(1) \otimes U_A(1)$
theories.
\end{quote}

\vfill

\begin{flushleft}
August 1995

\end{flushleft}
\end{titlepage}

\newpage

\renewcommand{\thefootnote}{\arabic{footnote}}
\setcounter{footnote}{0}
\newpage
\pagenumbering{arabic}

\section{Introduction}

Recently \cite{MGMW}, we have described the solution of the
 constraints
 for $(2,2)$ supergravity
in $(2,2)$ superspace  in terms of  unconstrained prepotentials.
For the
nonminimal $U_V(1)
\otimes U_A(1)$ theory, these are a  real vector superfield $H^m$
 and a
 general
scalar superfield $S$. In the ``degauged'' minimal theory where one
 of the
 $U(1)$ connections
is set to zero, a further constraint expresses the superfield $S$ in terms
 of a
chiral superfield $\s$ (for the $U_A(1)$ theory), or a twisted chiral
superfield
 $\tilde{\s}$
(for the $U_V(1)$ theory). We gave explicit expressions for the
vielbein, the
connections, and the vielbein determinant $E$. With these results
at hand,
 several
applications become possible. In particular, we have studied the
 theory in
light-cone gauge and derived the Ward identities associated with the
 (nonlocal)
induced $(2,2)$ supergravity action \cite{lc}. One can also
discuss the
fully supersymmetric quantization of the theory.  We have described
 in a
 separate work the general coupling of two-dimensional $(2,2)$
supersymmetric matter to supergravity  and the corresponding
 component
actions \cite{JGMGMW}.

For this latter application, a knowledge of invariant superspace
 measures
 in full
superspace and in chiral (or twisted chiral) superspace is
 necessary, as
 well as projection
formulae that allow one to obtain the corresponding component
actions.
In this paper we describe this aspect of the theory -- the
construction of
invariant actions, the determination of the measures (in particular
 of chiral
 or twisted
chiral densities), and the projection formulae that enable us to go from
 superspace actions
to component actions.
In the words of an esteemed colleague ``the construction of
superinvariants
 is  more an art
than a science". We show here that, at least in the two-dimensional
 $(2,2)$ theory,
this construction is in fact  a science.

Our paper is organized as follows: in section 2 we summarize the
 relevant information
about the $(2,2)$ supergravity theory and the solution of the
 constraints. In section 3
we show how, for the minimal $U_A(1)$ theory, one obtains the
 chiral measure for
integration over chiral superspace (or, equivalently, the twisted
chiral results for the
$U_V(1)$ theory). In section 4, we derive the component projection
 formula for
going from full superspace (or chiral superspace) to components.
 In the same section
 we
obtain an alternative projection formula which involves at an
intermediate
stage a twisted chiral projector rather than the chiral projector.
In section 5, we show
how to express the covariant derivatives of the $U_V(1) \otimes
 U_A(1)$ theory in terms
of those of the degauged  $U_A(1)$ theory, and derive the chiral
 measure for the
former.  Section 6 contains
our conclusions. We use the conventions of refs. \cite{MGMW,lc}.
 Appendix  A
contains the definition of the supergravity component fields in
 Wess-Zumino gauge,
and additional information about these derivatives. Appendices B
 and C contain some
details of the derivations in the main text.

\section{The (2,2) constraints and their solution}

We summarize in this section the main results of refs. \cite{MGMW,lc}.
 Tangent
space  Lorentz, $U_V(1)$ and $U_A(1)$  generators,
denoted here by ${\cal M}$, ${\cal Y}$ and
${\cal Y}'$ respectively, are defined by their  action on spinors:
\bea
[~{\cal M} \, , \, \psi_{\pm}~] ~=~ \pm \frac{1}{2}
 \psi_{\pm} ~~~~&,&~~~~
[~{\cal M} \, , \, \psi_{\pmd}~] ~=~ \pm \frac{1}{2}
\psi_{\pmd} ~~~,
\nonumber\\
{[}~{\cal Y} \, , \,  \psi_{\pm}~] ~=~  - \frac{i}{2}
\psi_{\pm} ~~~~&,&~~~~
[~ {\cal Y} \, , \, \psi_{\pmd}~] ~=~ +  \frac{i}{2}
\psi_{\pmd} ~~~,
\nonumber\\
{[}~ {\cal Y}' \, , \, \psi_{\pm}~] ~=~ \mp  \frac{i}{2}
\psi_{\pm}~~~~&,&~~~~
[~ {\cal Y}' \, , \, \psi_{\pmd}~ ]  ~=~ \pm  \frac{i}{2}
\psi_{\pmd} ~~~  .
\eea
It is also useful to define the combinations
\bea
M &=& \frac{1}{2}({\cal M} +i {\cal Y} ')~~~~,~~~
 \bar{M}= \frac{1}{2}({\cal M} -i {\cal Y} ') \non\\
N &=& \frac{1}{2}({\cal M} +i {\cal Y} )~~~~,~~~
 \bar{N}= \frac{1}{2}({\cal M} -i {\cal Y} )~~.
\eea
The covariant derivatives  are defined by
\bea
{\Del}_A  &=& E_A + \F_A {\cal M} + \S_A'  {\cal Y}'+ \S_A
 {\cal Y}  \non\\
          &=& E_A + \O_A M + \G_A \Mb + \S_A {\cal Y}  ~~.
\eea
They  satisfy the  constraints which define the 2D, N = 2 $U_V (1)
 \otimes U_A (1)$
supergravity,
\bea
 \{ { \nabla}_+ ~,~ { \nabla}_+ \} ~&=&~ 0 ~~~~~~~~~~~,~~~~~~
\{ { \nabla}_-
{}~,~ { \nabla}_- \} ~=~ 0 \nonumber\\
\{ {  \nabla}_+ ~,~ {  \nabla}_{\pd} \} ~&=&~ i  {  \nabla}_{\pp}
 ~~~~~~~,~~~~~~~ \{ {  \nabla}_- ~,~ { \nabla}_{\md} \} ~=~ i
{ \nabla}_{\mm} ~~~ \nonumber\\
\{ { \nabla}_+ ~,~ {  \nabla}_- \} ~&=&~ - \,  {\bar R} \,
{\bar M} ~~~~~,~~~~~ \{ { \nabla}_+ ~,~ { \nabla}_{\md} \} ~=~ - \,
{\bar F}  \, {\bar N} ~~~.
\eea
{}From the constraints follow the additional commutators
\bea
{[}~{ {\Del}}_{+} \, , \, { {\Del}}_{\pp} ~] &=& 0 ~~~, ~~~[~
{ {\Del}}_{-} \, , \, { {\Del}}_{\mm} ~] = 0
{}~~~, \\ \non
{[}~{ {\Del}}_{\pd} \, , \, { {\Del}}_{\pp} ~] &=& 0 ~~~, ~~~[~
{ {\Del}}_{\md} \, , \, { {\Del}}_{\mm} ~] =
0 ~~~, \\ \non
{[}~{ {\Del}}_{+} \, , \, { {\Del}}_{\mm} ~] &=& - \sihalf \Rb
{ {\Del}}_{\md}  -i({ {\Del}}_{\md} \Rb) \Mb
 - \sihalf \bar{F} {\Del}_-
-i ({ {\Del}}_-  \bar{F} ) \bar{N}
{}~~~, \\ \non
{[}~{ {\Del}}_{\pd} \, , \, { {\Del}}_{\mm} ~] &=&  \sihalf R
{ {\Del}}_{-}  +i( { {\Del}}_- R) M
+ \sihalf F
{ {\Del}}_{\md}  +i( { {\Del}}_{\md} F) N ~~~, \\ \non
{[}~{ {\Del}}_{-} \, , \, { {\Del}}_{\pp} ~] &=&  \sihalf \Rb
{ {\Del}}_{\pd}  -i({ {\Del}}_{\pd} \Rb)\Mb
 +\sihalf F
{ {\Del}}_{+}  -i({ {\Del}}_{+} F)N
{}~~~, \\ \non
{[}~{ {\Del}}_{\md} \, ,\,  { {\Del}}_{\pp} ~] &=& - \sihalf R
{ {\Del}}_{+}  +i({ {\Del}}_{+} R) M
-\sihalf \bar{F}
{ {\Del}}_{\pd}  +i({ {\Del}}_{\pd} \bar{F}) \bar{N}
 ~~~,
\eea
and
\newpage
\bea
{[}~ { {\Del}}_{\pp} \, , \, { {\Del}}_{\mm} ~] &=&  \half (
{ {\Del}}_+ R) { {\Del}}_- +\half ({ {\Del}}_-  R) { {
\Del}}_+  - \half ({ {\Del}}_{\pd} \Rb) { {\Del}}_{\md} - \half (
{ {\Del}}_{\md} \Rb) { {\Del}}_{\pd} \non \\
&& -  \half R \Rb \Mb - \half R \Rb M + ({{ {\Del}}}^2 R) M -(
{{ {\Delb}}}^2 \Rb) \Mb \non \\
&& + \half (
{ {\Del}}_+ F) { {\Del}}_{\md} +\half ({ {\Del}}_{\md} F)
 { {
\Del}}_+  - \half ({ {\Del}}_{\pd} \bar{F}) { {\Del}}_{-}
 - \half (
{ {\Del}}_{-} \bar{F}) { {\Del}}_{\pd} \non \\
&& -  \half F \bar{F} \bar{N} - \half F \bar{F} N +
({ {\Del}}_+ { {\Del}}_{\md} F) N -(
{ {\Del}}_{\pd} { {\Del}}_-  \bar{F}) \bar{N}
\label{DelppDelmm} ~~.
\eea
Furthermore,
for the minimal supergravities one restricts the gauge group
so that either
$F=0$ for the $U_A(1)$ version,
or $R=0$ for the $U_V(1)$ version, by  ``degauging'',  i.e. by
 setting either
$\Sigma_{A}=0$ or
${\S}_{A}'=0$.

The solution of the constraints is obtained in terms of the  ``hat''
differential operators
\beq
\hat{E}_{\pm}= e^{-H}D_{\pm}e^{H}~~~~,~~~~ \hat{E}_{\pmd}
= e^{H} D_{\pmd}
 e^{-H}
\eeq
with $H=H^mi\pa_m$,  where $H^m$ is a real vector superfield.
The spinorial vielbein is expressed in terms of these operators,
 as well as an
additional (superscale and $U(1)$  compensator) general complex
superfield $S$
as
\bea
E_+ \equiv e^{\Sb}(\Eh _+ + \Apm \Eh_-) ~~~~&,&~~~~E_- \equiv
 e^{\Sb}
(\Eh_-+\Amp \Eh_+) \non\\
E_{\pd} \equiv e^S(\Eh _{\pd} + \Apmd \Eh_\md)
 ~~~~&,&~~~~E_\md
\equiv  e^S
(\Eh_\md+\Ampd
\Eh_\pd) ~~.
\eea
The vectorial vielbein is calculated from the constraints.
The $A$'s, as well as the vielbein determinant $E$ are given explicitly
 in ref. \cite{MGMW}.
 We emphasize that they are functions  of the superfield  $H^m$
 {\em only}.

The connections were derived in ref.  \cite{MGMW}, and we list them
 here for convenience:
\bea
{\Omega}_+&=&+ e^{{\Sb}}(\Eh_-{\Apm} - {\Apm}\Eh_+ {\Amp} )\non\\
{\Omega}_- &=&-e^{{\Sb}}( \Eh_+{\Amp} -{\Amp} \Eh_-{\Apm})  \non\\
{\Sigma}_+&=&-2ie^{{\Sb}}(\Eh_+{\Sb }+{\Apm}\Eh_-{\Sb} )-ie^{{\Sb}}
(\Eh_-{\Apm } -{\Apm}\Eh_+{\Amp} ) \label{connections} \\
{\Sigma}_-&=&-2ie^{{\Sb}}(\Eh_-{\Sb} +{\Amp} \Eh_+{\Sb} )-ie^{{\Sb}}
(\Eh_+{\Amp} -{\Amp}\Eh_-{\Apm} ) \nonumber\\
{\Gamma}_+ &=& +2e^{{\Sb}}(\Eh_+S+{\Apm }\Eh_-{S})+2e^{{\Sb}}
(\Eh_+{\Sb}
 +{\Apm} \Eh_-{\Sb})+e^{{\Sb}}(\Eh_-{\Apm} -{\Apm} \Eh_+{\Amp} ) \non\\
{\Gamma}_- &=& -2e^{{\Sb}}(\Eh_-{S}+{\Amp} \Eh_+{S}-2e^{{\Sb}}
(\Eh_-{\Sb}
 +{\Amp} \Eh_+{\Sb})-e^{{\Sb}}(\Eh_+{\Amp} -{\Amp} \Eh_-{\Apm }) \non
\eea

The   $U_V(1) \otimes U_A(1)$ theory contains in addition to the
 irreducible
supergravity multiplet a vector multiplet. A minimal theory is obtained by
 setting one
of the field strengths $R$ or $F$  (or the corresponding connections)
 to zero.
 In ref. \cite{MGMW} we
worked out the implications of the additional constraint $F=0$ for
 the minimal
 $U_A(1)$
theory and found that the superfield $S$ satisfies a constraint that
expresses
 it
in terms of an arbitrary covariantly chiral superfield $\s$ as follows:
\beq
e^S = e^{{\s}} \frac{ \left[1\cdot e^{\BAH}
 \right]^{-\frac{1}{2}}}{[1-\Apmd \Ampd]^{\frac{1}{2}}}
 E^{-\frac{1}{2}}~~~~,
{}~~~~e^{\Sb} = e^{\bar{\s}} \frac{ \left[1\cdot e^{-\BAH}
 \right]^{-\frac{1}{2}}}{[1-\Apm \Amp]^{\frac{1}{2}}} E^{-\frac{1}{2}}~~,
 \label{degauge}
\eeq
where  $\BAH$ indicates
that the differential operator in $H^m i \pa_m$ acts on objects
 to its left.
The unconstrained real vector superfield $H^m$ and the chiral
scalar
superfield $\s$ are the  prepotentials of  the minimal
$U_A(1)$ $(2,2)$
supergravity. The mirror image $U_V(1)$ theory is obtained
by interchanging
$-$ and $\md ~$ indices everywhere (and interchanging $R$
 with $F$, and
 $M$ with $N$), and amounts
to replacing the chiral superfield $\s$ by a twisted chiral superfield
 $\tilde{\s}$.
In the following, we shall concentrate primarily on the minimal
 $U_A(1)$ theory.

\sect{From full superspace measure to chiral superspace measure}

In full superspace, invariant actions are constructed by means of the
 vielbein
determinant as
\beq
{\cal S} = \int d^2x d^4 \th E^{-1} {\cal L}
\eeq
where ${\cal L}$ is an arbitrary scalar function of superfields. The
 invariance
 of
this integral under superspace coordinate transformations can be
 established by
standard means; see for example {\em Superspace}, sect. 5.5
cite{Superspace}.
In addition \cite{Superspace,lc}, one can show that for superspace
 transformations under which
covariantly chiral and antichiral superfields transform as
\beq
\Phi \rightarrow  e^{i\L}\Phi ~~~~,~~~ \bar{\Phi} \rightarrow
 e^{i \bar{\L}}
\bar{\Phi}
\eeq
with
\bea
\L&=&\L^m i\pa_m +\L^{\a}iD_{\a} +\L^{\ad}iD_{\ad} \non\\
\bar{\L}&=&\bar{\L}^m i\pa_m +\bar{\L}^{\a}iD_{\a} +
\bar{\L}^{\ad}iD_{\ad} ~~,
\eea
the  prepotential  $H^a$  must transform as
\beq
e^{2H} \rightarrow e^{i\bar{\L}} e^{2H} e^{-i\L}
\eeq
and the vielbein determinant transforms as
\beq
E^{-1} e^{\BAH} \rightarrow  E^{-1}e^{\BAH} e^{i \buildrel
\leftarrow \over
{\bar{\L}}} ~~.
\eeq

In the minimal $U_A(1)$ theory we  show now  that  one  can
 rewrite the full superspace integral as an integral over chiral
superspace, with an appropriate measure  ${\cal E}$ that we
determine
 explicitly:
\beq
{\cal S}= \int d^2x d^2 \th {\cal E}^{-1} \Delb^2 {\cal L}|_{\bar{\th}=0}
\eeq
with
\beq
{\cal E}^{-1} =  e^{-2\s} (1.e^{\BAH}) ~~.
\eeq

We proceed by deriving first an explicit expression for  $\Delb^2$
 acting
on an {\em arbitrary} scalar  $L$.
 We start with
\bea
\Delbsq L&=& \Delpd \Delmd L \non \\
&=& [e^S(\Eh_\pd + \Apmd \Eh_\md) - \half \O_\pd]
     [e^S(\Eh_\md + \Ampd \Eh_\pd)] L  ~~.
\eea
{} From the complex conjugate of equations \reff{connections},
 with $\S_\pd
 = 0$,
we have
\beq
\O_\pd = -2e^S(\Eh_\pd S + \Apmd \Eh_\md S) ~~.
\eeq
Using this we  rewrite $\Delbsq L$ as
\bea
\Delbsq L &=& (\Eh_\pd + \Apmd \Eh_\md)
     e^{2S}(\Eh_\md + \Ampd \Eh_\pd) L   ~\nonumber\\
&=&{(\Eh_\pd + \Apmd \Eh_\md) (\Eh_\md + \Ampd
 \Eh_\pd) e^{2S} L} \non \\
&&-2 (\Eh_\pd + \Apmd \Eh_\md)(\Eh_\md S + \Ampd \Eh_\pd  S)
e^{2S} L  ~~.
\eea
Again from the complex conjugate of  equations \reff{connections},
 setting
 $\S_\md = 0$, we obtain
\beq
2(\Eh_\md S + \Ampd \Eh_\pd  S) = -(\Eh_\pd \Ampd - \Ampd
\Eh_\md \Apmd) ~~,
\eeq
which, when substituted into the expression for $\Delbsq L$ above,
gives
\bea
\Delbsq L&=& [(1-\Apmd \Ampd)\Eh_\pd \Eh_\md + \Eh_\md(\Apmd
  \Ampd)\Eh_\pd - \Eh_\pd(\Apmd \Ampd)\Eh_\md \non \\
&& - \Eh_\pd \Eh_\md (\Apmd \Ampd)] e^{2S} L  \non\\
&=& \Eh_\pd \Eh_\md [(1-\Apmd \Ampd)e^{2S} L]~~.
\eea
We have used the complex conjugates of (3.5) and (3.6) of ref.
 \cite{MGMW}, namely,
\bea
\Eh_\pd \Apmd + \Apmd \Eh_\md \Apmd &=& 0 \non \\
\Eh_\md \Ampd + \Ampd \Eh_\pd \Ampd &=& 0
\eea
and
\bea
\Eh_\pd \Eh_\md \Apmd &=& 0  \non \\
\Eh_\md \Eh_\pd \Ampd &=& 0  ~~.
\eea
We obtain, using  \reff{degauge}
\beq
\Delbsq L  = \Eh_\pd \Eh_\md [e^{2 \s} (1.e^{\BAH})^{-1}E^{-1} L] ~~.
 \label{ident}
\eeq

  We define $ L'$ by   $L = e^{-2 \s}(1.e^{\BAH}) L'$,  so that
\reff{ident} becomes
\bea
 \Delbsq [e^{-2 \s}(1.e^{\BAH}) L']
&=& \Eh_\pd \Eh_\md (E^{-1} L') \non \\
&=& e^H \bar{D}^2 e^{-H}(E^{-1} L')  ~~.
\eea
Multiplying  both sides by $e^{-H}$ and integrating over chiral
 superspace, we
 obtain
\beq
\int  d^2x d^2 \th \bar{D}^2 e^{-H} E^{-1} L' =\int d^2x d^2 \th e^{-H}
e^{-2 \s} (1.e^{\BAH})L' ~~.
\eeq
Integrating the operator  $e^{-H}$ by parts on both sides
of the equation
 gives
\beq
\int d^2x d^2 \th  \bar{D}^2 [(1.e^{\BAH})
  E^{-1} L']  = \int d^2x d^2 \th
 (1.e^{\BAH}) \Delbsq [e^{-2 \s}(1.e^{\BAH}) L']  ~~.
\eeq
Finally, defining ${\cal L}$ by  $ L'    = (1.e^{\BAH})^{-1}  {\cal L}$, we
 obtain
\beq
\int d^2x d^2 \th \bar{D}^2 [E^{-1} {\cal L}] =\int d^2x d^2 \th
 e^{-2 \s}(1.e^{\BAH}) \Delbsq {\cal L} ~~.
\eeq
We have used the fact that $\s$ is covariantly chiral to pull the factor
$e^{-2 \s}$ past the covariant derivatives.
Since one can go from a full superspace integral to a $d^2 \th$
 integral by
\beq
\int d^2x d^4 \th E^{-1} {\cal L} = \int d^2x d^2 \th  \bar{D}^2  [E^{-1}
 {\cal L}] |_{\bar{\th}
=0}
\eeq
the relation above leads us to the desired result (and determines the
 chiral
measure ${\cal E}$):
\beq
\int d^2x d^4 \th E^{-1} {\cal L} = \int d^2 x d^2 \th  e^{-2\s}
 (1.e^{\BAH})
\bar{\Del}^2 {\cal L} |_{\bar{\th}=0}
 ~~.   \label{chint}
\eeq

By construction, the above integral is an invariant. We can replace
 $\bar{\Del}^2 {\cal L} $
by any covariantly chiral expression ${\cal L}_{chiral}$,  transforming
 under
 superspace coordinate
transformations as
\beq
{\cal L}_{chiral} \rightarrow e^{i {\L}}{\cal L}_{chiral}
\eeq
and deduce the transformation properties of the measure
\beq
e^{-2\s} \left( 1\cdot e^{\BAH} \right)\rightarrow e^{-2\s}
\left( 1\cdot e^{\BAH}
\right)
e^{i \buildrel \leftarrow \over \L} ~~.
\eeq
Obviously, a similar result holds for the decomposition to an
antichiral
 integral.

In the $U_V(1)$ theory corresponding results, in terms of the
 twisted chiral
compensator, can be obtained for  twisted chiral integrals of the type
$
\int d^2x d \th^+ d\th^{\md}
$.

\sect{From  full superspace measure to component integrals}

\subsection{The  chiral density projection formula}

We have shown in the previous section that it is possible to
express the full
superspace integral in terms of a chiral integral. In the next step of the
 projection
we rewrite this as a component integral over ordinary  $d^2x$
 space.
By dimensional arguments, and from an examination of the index
structure of
 the possible
terms,  the density formula for a general $\cal L$ must take the form
\bea
 \int d^2 x d^4 \th E^{-1} {\cal L} &=& \int d^2x d^2 \th {\cal E}^{-1}
 \bar{\Del}^2{\cal L}
|_{\bar{\th}=0}
 \non \\
&=& \int d^2 x e^{-1} [\Del^2 + X^+ \Del_+ + X^- \Del_- +Y]
\bar{\Del}^2
{\cal L} |_{\th = \bar{\th}=0}
     ~~ , \label{gendens}
\eea
where $e$ is the ordinary space zweibein determinant and
 $X^+$, $X^-$, and
 $Y$ are to be determined. In the past, the determination of the
coefficients $X$, $Y$, has been done  essentially by requiring
 that the final
 component action
be invariant under supersymmetry transformations.
Our  derivation is  based on the idea that we should obtain the
 same result
 for the component
action whether  we go through the intermediate stage of a chiral
 integral,
  as in the
equation above, or through a corresponding
antichiral integral.
We illustrate the procedure  by using for ${\cal L}$ the
free lagrangian for the chiral multiplet.
As we shall see, proceeding from the above expression, we
obtain a result which is not  automatically symmetric in the auxiliary
fields
 $F$, $\bar{F}$
of the chiral multiplet. Requiring that the result be symmetric leads to a
 unique determination
of the coefficients $X$, $Y$.

We consider the  kinetic action  for the chiral multiplet
\bea
\int d^2 x d^4 \th E^{-1} \bar{\Phi}\Phi
&=& \int d^2 x e^{-1} [\Delsq + X ^\a \Del_\a  + Y] \Delbsq
 (\bar{\F}\F)|     \non \\
&=& \int d^2 x e^{-1}[(\Delsq \Delbsq \bar{\F}) \F| + (\Delp \Delbsq
 \bar{\F})
 (\Delm \F)|
 - (\Delm \Delbsq \bar{\F})(\Delp \F)| \non \\
&& + (\Delbsq \bar{\F}) (\Delsq \F)| + X^+ \Delp (\Delbsq \bar{\F} \F)
 |\non \\
&& + X^- \Delm (\Delbsq \bar{\F} \F) |+ Y(\Delbsq \bar{\F}) \F|]
 \label{kinetic}
\eea
according to the density formula.

The (covariant) components of the chiral multiplet  are defined by
\bea
\F| = \f &,& \bar{\F}| = \bar{\f}  \nonumber\\
\Delp \F | = \psi _+ &,& \Delpd \bar{\F}| = \psi_\pd  \nonumber \\
\Delm \F | = \psi_- &,& \Delmd \bar{\F}| = \psi_\md \nonumber \\
\Del^2\F| = -iF &,&  \Delb^2\bar{\F}|
=-i \bar{F}
.\eea
 supergravity
Appendix A.
In Appendix B we present the complete expressions for the other
component
quantities  (involving additional  derivatives of $\Phi$)
 appearing above.
For our present purposes, we find that the unknowns can be
 determined just by
looking at the terms in those expressions that contain the auxiliary
fields, $F$, $\bar{F}$.
 We list the relevant terms here:
\bea
\Delp \Delbsq \bar{\F}| &\sim&   \psi_\pp^\pd \bar{F}\\
\Delm \Delbsq \bar{\F}| &\sim&   \psi_\mm^\md \bar{F}\\
\Delsq \Delbsq \bar{\F}| &\sim&
 -i( \psi_\pp^\md \psi_\mm^\pd \bar{F} - \psi_\pp^\pd \psi_\mm^\md
\bar{F})
-\frac{i}{2} \bar{B} \bar{F} ~~,
\eea
where   we  have also used the definition $R| = B $, $ \Rb| = \bar{B}$ .

Substituting into  \reff{kinetic},  the sum of the terms
containing auxiliary fields is
\bea
&&i \psi_\pp^\pd \psi_\mm^\md \bar{F} \f -i \psi_\pp^\md
\psi_\mm^\pd
\bar{F} \f
+ \psi_\pp^\pd \bar{F} \psi_- - \psi_\mm^\md \bar{F} \psi_+ \non \\
&-& \sihalf \bar{B} \bar{F} \f - \bar{F} F +  X^+ \psi_\pp^\pd \bar{F} \f
-i X^+ \bar{F} \psi_+ -i X^- \bar{F} \psi_- \non \\
&+&  X^- \psi_\mm^\md \bar{F} \f -i Y \bar{F} \f ~~.
\eea
Clearly, except for the $\bar{F} F$ term, this expression is not
symmetric in
barred and unbarred
quantities,  and will not agree with the result  we would obtain
 by going
 through
the intermediate stage of an antichiral integral unless we set to
zero the
 coefficients of the $\bar{F}\f
$ and the
$\bar{F}\psi_{\pm}$
terms. We obtain
\beq
X^+ = i \psi_\mm^\md ~~,~~ X^- = -i \psi_\pp^\pd ~~,~~
Y = - \half \bar{B} - \psi_\pp^\md \psi_\mm^\pd + \psi_\mm^\md
\psi_\pp^\pd~~,
\eeq
leaving $\bar{F} F$ as the only contribution. The form of the chiral
 density
formula is
thus determined.

We have, for the final result,
 the chiral density projection formula
\bea
&& \int d^2 x d^4 \th E^{-1} {\cal L}  \non \\
&=&  \int d^2 x e^{-1} [ \Del^2 +i \psi_=^{\dot{-}}\Del_+ -i
\psi_{\pp}^{\pd}\Del_-
+(- \half \bar{B} -\psi_{\pp}^{\dot{-}}\psi_=^{\pd}
+\psi_=^{\dot{-}}\psi_{\pp}^{\pd})]\bar{\Del}^2
{\cal L} | ~~.  \label{chiral density}
\eea
This component expression can be rewritten in terms of a twisted
 chiral
projector  as discussed in the next subsection.

\subsection{The twisted chiral density projection formula}
In  the previous section we have written down the density formula
 for going
 from a
full superspace integral to a component expression containing as
an intermediate
ingredient a chiral integrand $\Delb ^2 {\cal L}$. In this section
we will
 derive a
similar formula involving a {\em twisted} chiral integrand
$\Del_{\pd} \Del_-
 {\cal L}$,
\bea
&& \int d^2 x d^4 \th E^{-1} {\cal L}  \non \\
&=&\int d^2 x  e^{-1}[ \Del_{\md}\Del_+ +i \psi^{\pd}_{\pp}
\Del_{\md}
-i\psi^{-}_{\mm} \Del_{+} +(\psi^{\pd}_{\pp}\psi^-_{\mm}
+\psi^-_{\pp}\psi^{\pd}_{\mm})]
\Del_{\pd}\Del_{-} {\cal L}|
{}~.
\eea

We start  with the following general identity,
derived  straightforwardly from
the commutation relations and the Bianchi identities
\bea
\lefteqn{\Del^2 \bar{\Del}^2 +\bar{\Del}^2\Del^2 +\Del_-
\bar{\Del}^2
\Del_+
+\Del_{\md}\Del^2\Del_{\pd}} \nonumber\\
&=& \Del_{\pp}\Del_{\mm} -\half \Del_+R \Del_- + \half
\Del_{\pd}\bar{R}\Del_{\md}
\nonumber\\
&=&\half \Del_{\pp}\Del_{\mm} +\half \Del_{\mm}\Del_{\pp}
+ {\frac{1}{4}}[- (\Del_+R)\Del_-  + (\Del_-R)\Del_+ +(\Del_{\pd}
\bar{R})
\Del_{\md}
-(\Del_{\md}\bar{R})\Del_{\pd}]\nonumber\\
&&- \half [R\Del^2 -\bar{R}\bar{\Del}^2] ~~. \label{genident}
\eea
We apply this sum of operators to an arbitrary scalar (so that
certain
connection terms
can be dropped), inside a $d^2x e^{-1}$ integral (so that
space-time
 derivatives can
be dropped),
 and then evaluate  at $\th =0$, using some of the component
expressions for
covariant
derivatives from Appendix A, etc.

After some lengthy manipulations, that we outline in Appendix C,
we obtain
the following identity:
\bea
\int d^2 x e^{-1}
\lefteqn{\{\Del^2 \bar{\Del}^2 +\bar{\Del}^2\Del^2 +\Del_-
 \bar{\Del}^2\Del_+
+\Del_{\md}\Del^2\Del_{\pd}  \}{\cal L}| }  \nonumber\\
&=& \int d^2x e^{-1} \{
[\psi_{\pp}^+ \psi_\mm^- +\psi_{\pp}^- \psi_\mm^+ - \half B ]
 \Del^2
+[\psi_{\pp}^{\pd}\psi_\mm^{\md}+\psi_{\pp}^{\md}\psi_\mm^{\pd}
+\half
\bar{B}]\bar{\Del}^2 \nonumber\\
&+&[\psi_{\pp}^+\psi_\mm^{\md}+\psi_{\pp}^{\md}\psi_\mm^+]
\Del_+\Del_{\md}
+[\psi_{\pp}^{\pd}\psi_\mm^- +\psi_{\pp}^-\psi_\mm^{\pd}]
\Del_{\pd}
\Del_- \nonumber\\
&+&i\psi_{\pp}^+[\Del_{\md}\Del^2 +\Del_-(\Del_+\Del_{\md})]
-i\psi_\mm^-[\Del_{\pd}\Del^2 +\Del_+(\Del_{\pd}\Del_-)] \non\\
&+&i\psi_{\pp}^{\pd}[\Del_- \bar{\Del}^2 +\Del_{\md}
(\Del_{\pd}\Del_-)]
-i\psi_\mm^{\md}[\Del_+\bar{\Del}^2 +\Del_{\pd}
(\Del_+\Del_{\md})] \} {\cal L}|
{}~~.
\eea
On the left hand side of the equation we have the sum of projectors.

We move now certain terms from one side of the equation to the other and
 obtain
\bea
&&\int d^2 x  e^{-1}[\Del^2 -i\psi^{\pd}_{\pp}\Del_- +i
\psi^{\md}_{\mm}
\Del_+
-(\psi_{\pp}^{\pd}\psi_\mm^{\md}+\psi_{\pp}^{\md}
\psi_\mm^{\pd} +\half
\bar{B})]\Delb^2 {\cal L}| \nonumber\\
&&+\int d^2 x  e^{-1}[\Delb^2 -i \psi^+_{\pp}\Del_{\md} +i
\psi^-_{\mm}\Del_{\pd}
-(\psi_{\pp}^+ \psi_\mm^- +\psi_{\pp}^- \psi_\mm^+ - \half B)]
\Del^2 {\cal L}|
\nonumber\\
&& = \int d^2 x  e^{-1}[ \Del_{\md}\Del_+ +i \psi^{\pd}_{\pp}
\Del_{\md}
-i\psi^{-}_{\mm} \Del_{+} +(\psi^{\pd}_{\pp}\psi^-_{\mm}
+\psi^-_{\pp}\psi^{\pd}_{\mm})]
\Del_{\pd}\Del_{-} {\cal L}| \nonumber\\
&&+ \int d^2 x  e^{-1} [\Del_-\Del_{\pd} + i\psi^+_{\pp} \Del_{-}- i
\psi^{\md}_{\mm} \Del_{\pd}
+(\psi^+_{\pp}\psi^{\md}_{\mm} +\psi^{\md}_{\pp}\psi^+_{\mm})]
\Del_+\Del_{\md}
 {\cal L}|
\eea
However, it can be shown that up to total derivatives the two
 terms on each
side of the
equation are equal. We obtain therefore the following result,
which allows us
to
write a full superspace integral in terms of either chiral or twisted
 chiral
projections,
\bea
&& \int d^2 x d^4 \th E^{-1} {\cal L} \label{twistdensity}\\
&&= \int d^2 x e^{-1} [\Del^2 -i\psi^{\pd}_{\pp}\Del_- +i
\psi^{\md}_{\mm}\Del_+
-(\psi_{\pp}^{\pd}\psi_\mm^{\md}+\psi_{\pp}^{\md}\psi_\mm^{\pd}
 +\half
\bar{B})]\Delb^2 {\cal L}|\nonumber\\
&&=\int d^2 x  e^{-1}[ \Del_{\md}\Del_+ +i \psi^{\pd}_{\pp}
\Del_{\md}
-i\psi^{-}_{\mm} \Del_{+} +(\psi^{\pd}_{\pp}\psi^-_{\mm}
+\psi^-_{\pp}\psi^{\pd}_{\mm})]
\Del_{\pd}\Del_{-} {\cal L}|  ~~.\non
\eea
(The asymmetry between the two forms is due to the constraint
$F=0$.)
Replacing $\Delbsq {\cal L}$ or $\Delpd \Delm {\cal L}$ by
 arbitrary chiral
 or twisted
chiral lagrangians gives the projection formulae for chiral or
 twisted chiral
 actions,
respectively.

\sect{From full superspace measure to chiral superspace
 measure in the
$U_V(1) \otimes U_A(1) $ case}

   In this section we derive the analogue of \reff{chint} for the
undegauged
 $U_V(1) \otimes
U_A(1)$ supergravity theory.   We accomplish this by relating
 $\Delbsq$ in
 the degauged case to ${\Delbsq}$ in the $U_V(1) \otimes
 U_A(1)$ case.
 We consider first the covariant derivative
\beq
{\Delpd} = E_\pd + {\Omega}_\pd \Mb + {\G}_\pd M +
{\S}_\pd {\cal Y} ~~,
\eeq
and substitute the explicit expressions for the connections:
\bea
{\Omega}_\pd &=&+ e^{{S}}(\Eh_\md {\Apdmd} - {\Apdmd }
\Eh_\pd {\Amdpd} )
\nonumber\\
{\Omega}_\md &=&-e^{{S}}( \Eh_\pd {\Amdpd}  - {\Amdpd}
\Eh_\md {\Apdmd}) \non\\
{\Sigma}_\pd &=& 2ie^{{S}}(\Eh_\pd {S}+{\Apdmd}\Eh_\md {S} )
+ie^{{S}}(\Eh_\md
 {\Apdmd } -{\Apdmd}
\Eh_\pd {\Amdpd} ) \non\\
{\Sigma}_\md &=& 2ie^{{S}}(\Eh_\md {S} +{\Amdpd} \Eh_\pd {S} )
+ie^{{S}}
(\Eh_\pd {\Amdpd} -{\Amdpd}
\Eh_\md {\Apdmd} ) \nonumber\\
{\Gamma}_\pd &=& +2e^{{S}}(\Eh_\pd {\Sb} + {\Apdmd }
\Eh_\md {\Sb})
+2e^{{\Sb}}
(\Eh_\pd {S} +{\Apdmd} \Eh_\md {S}
)+e^{{S}}(\Eh_\md {\Apdmd} -{\Apdmd} \Eh_\pd {\Amdpd} ) \non\\
{\Gamma}_\md &=& -2e^{{S}}(\Eh_\md {\Sb}+{\Amdpd}
\Eh_\pd {\Sb}-2e^{{\Sb}}
(\Eh_\md {S} +{\Amdpd} \Eh_\pd {S}
)-e^{{S}}(\Eh_\pd {\Amdpd} -{\Amdpd} \Eh_\md {\Apdmd}) \non
\eea

We find
\beq
{\Delpd }= E_\pd + {\Omega}_\pd ({\cal M} + i {\cal Y}) +
 2E_\pd (S+ \Sb) M
+ 2i(E_\pd S) {\cal Y} ~~. \label{intdelpd}
\eeq
   We note the following identities
\bea
e^{-u{\cal Y}} \Eh_{\dot{\pm}} e^{u{\cal Y}}
&=& e^{-\frac{i}{2} u} \Eh_{\dot{\pm}}  + e^ {-\frac{i}{2}u}
(\Eh_{\dot{\pm}}u){\cal Y} ~~, \non\\
e^{-vM} \Eh_{\dot{\pm}} e^{vM}
&=&  \Eh_{\dot{\pm}}  + (\Eh_{\dot{\pm}}v)M ~~.\label{expident}
\eea
We obtain then, from \reff{intdelpd}
\bea
\lefteqn{e^{2({S}+{\Sb})M +2i {S}{\cal Y}} {\Delpd} e^{-2i{S}
{\cal Y} -2({S}
+{\Sb})M}=} \non\\
&&(\Eh_\pd + {\Apdmd} \Eh_\md) + (\Eh_\md {\Apdmd} -
{\Apdmd} \Eh_\pd
{\Amdpd)}({\cal M}+i{\cal Y}) ~~. \label{relation}
\eea

  We note now that \reff{relation} is valid in both the undegauged
 and degauged
theories and the right-hand-side is independent of $S$,
which is the only
 quantity that is
affected by the actual degauging.
Therefore the left-hand-sides for the $U_ V(1) \otimes U_A(1)$
 and the
 $U_A(1)$ theories can be set equal to each other and this
 yields a relation
 between the covariant
derivatives,
\bea
\lefteqn{\left( e^{2({\bf S}+\bar{\bf S})M +2i {\bf S}{\cal Y}} {\Delpd}
 e^{-2i{\bf S}{\cal Y} -2({\bf S}+\bar{\bf S})M}\right)
_{U_V(1) \otimes U_A(1)}
=} \non\\
&&\left( e^{2({S}+{\Sb})M +2i {S}{\cal Y}} {\Delpd} e^{-2i{S}
{\cal Y} -2({S}
+{\Sb})M}\right)
_{U_A(1)}
\eea
with a similar expression for $\Delmd$. Consequently
\beq
\left({\Del_\pmd}\right)
_{U_V(1) \otimes U_A(1)}
= e^{2({S}+{\Sb}-{\bf S} - \bar{\bf S})M +2i ({S - {\bf S}})
{\cal Y}} \left(
 {\Del_\pmd}\right)_{U_A(1)}
e^{-2i({S -{\bf S}}){\cal Y} -2({S}+{\Sb} -{\bf S} - \bar{\bf S})M}
 ~~,
\eeq
with a similar relation for the undotted derivatives.  To distinguish
 between
 the two cases, we
have denoted by
${\bf S}$ and $S$  the scale compensators in the $U_V(1) \otimes
 U_A(1)$
 and the $U_A(1)$
theories, respectively (the former is a general scalar superfield
whereas the
 latter is given by
\reff{degauge}).

  The above relation expresses the covariant derivatives of the
 $U_V(1)
 \otimes U_A(1)$ theory
in terms of the covariant derivatives of the $U_A(1)$ theory.
In particular
 we can write
\beq
\left(\Delbsq \right)_{U_A(1)}
= e^{-2({S}+{\Sb}-{\bf S} - \bar{\bf S})M -2i ({S - {\bf S}}){\cal Y}}
 \left({\Delbsq}
\right)_{U_V(1) \otimes U_A(1)}  e^{2i({S -{\bf S}}){\cal Y}
+2({S}+{\Sb}
 -{\bf S} - \bar{\bf S})M}~~.
\eeq
Acting on a scalar lagrangian ${\cal L}$ we can drop the
 exponentials
 following $\Delbsq$, and  acting on $\Delbsq {\cal L}$
 we can drop the
 $e^{(\cdots) M}$ term, while
\beq
e^{-2i( S- {\bf S}){\cal Y}} \Delbsq_{U \otimes U}  {\cal L}
 = e^{2(S-{\bf S})} \Delbsq_{U \otimes U} {\cal L} ~~.
\eeq
  Substituting  this into \reff{chint} we obtain
\bea
\int d^2x d^4 \th E^{-1} {\cal L}
&=& \int d^2 x d^2 \th  e^{-2\s} (1.e^{\BAH})\left({\bar{\Del}^2 }
\right)_
{U_A(1)}{\cal L} |_{\bar{\th}=0} \non\\
&=& \int d^2 x d^2 \th E^{-1} \frac{e^{-2{\bf S} }}{[1-{\Apdmd}
 {\Apdmd}]}
{{\bar{\Del}}^2}_{U \otimes U} {\cal L} |_{\bar{\th}=0}
 ~~,   \label{undegaugedchint}
\eea
which defines the chiral measure in the $U_V(1) \otimes U_A(1)$
 theory,
 and allows us to
construct invariant actions of the form $\int d^2x d^2 \th {\cal E}^{-1}
 {\cal L}_{chiral}$.

    Obviously, since the theory is symmetric under the
interchange of $-$
 and $\md$, a twisted
chiral measure exists for writing invariants $\int d^2x d^2 \tilde{\th}
 \tilde{{\cal E}}^{-1} {\cal L}_{twisted~chiral}$.  However,
as we discuss
 in the concluding section, an explicit
expression is harder to come by.

\sect{Discussion}

In this paper we have presented results for the integration
measures in the
 $(2,2)$ theories,
and have derived  projection formulae for  obtaining component
actions from the
corresponding superspace actions. This particular issue has been
 a sore point in
most derivations because no clear, completely superspace
 technique
exists in the general case. For the case at hand, at least, we have
 managed
 to avoid
any reference to component supersymmetry transformations
 \cite{Superspace}
or explicit $\th$ expansions of the supergravity prepotentials
in Wess-Zumino
 gauge
\cite{Buchbinder}.

Whereas in the degauged theory, which contains a chiral
compensator, the
 existence and
construction of the chiral measure is straightforward, this is
 not obviously
 the case in
the undegauged theory. We have shown, by explicitly
expressing the covariant
  derivatives
of the  $U_V(1) \otimes U_A(1)$ theory in terms of those
of the degauged,
 $U_A(1)$ theory,
how to obtain the chiral measure for the former theory,
from a knowledge of
 the chiral measure for the latter theory.

We have also argued that, by symmetry, a twisted chiral
measure exists. However,
its explicit construction is not straightforward. The point is
 that in our
 solution of the
constraints in ref. \cite{MGMW},  we have chosen to express
 the covariant
 derivatives in terms of the
objects $\Eh$  and the compensators $S$ which break the
 symmetry between
 $-$ and $\md$ objects. To maintain the symmetry we would
have to introduce
additional prepotentials which were eliminated from the
beginning by an
appropriate  $K$-transformation gauge choice.

The lack of symmetry manifests itself also in the construction
 of covariantly
 chiral
and covariantly twisted chiral superfields in terms of ordinary
 chiral and
 twisted
chiral ones.  The construction of covariantly chiral superfields is
straightforward,
\beq
\Phi_{cov.~chiral} = e^H \phi_{chiral}  \label{covchi}~~,
\eeq
because then the left-hand-side is annihilated by
$\Del_{\dot{\pm}}$ when the
right-hand-side is annihilated by  $D_{\dot{\pm}}$ (c.f. (2.7)).
 However, if we want
  to obtain a
 covariantly twisted chiral superfield ${\cal X}$ satisfying
\beq
\Del_{\pd} {\cal X} = \Del_- {\cal X} =0
\eeq
in terms of an ordinary twisted chiral superfield satisfying
\beq
D_{\pd} \chi = D_- \chi =0 ~~,
\eeq
it is clear that the above construction will not work. In fact,
 we have not
 succeeded
in obtaining a closed form relation similar to \reff{covchi}.  In the
 remainder of this
concluding section we show  what the relation is to first order in the
 supergravity
prepotential.

 We begin by writing down the differential equations that the twisted
chirality conditions imply for $\calX$,
\bea
{[} e^{-H}D_-e^H + \Amp e^{-H} D_+ e^H ] \calX &=& 0 \non \\
{[} e^H D_\pd e^{-H} + \Apmd e^H D_\md e^{-H}] \calX
 &=& 0  ~~.
\eea
To first order in $H$, these equations reduce to
\bea
D_- \calX + i (D_-H^a) \di_a \calX + \Amp D_+ \calX &=& 0 \non \\
D_\pd \calX - i (D_\pd H^a) \di_a \calX + \Apmd D_\md \calX
 &=& 0 ~~,
\eea
with
\bea
\Amp &=& -2 D_\pd D_- H^\pp + {\cal O}(H^2) \non\\
\Apmd &=& -2 D_\pd D_- H^\mm + {\cal O}(H^2)~~.
\eea
We now set $\calX = \chi + Z$, and solve iteratively for $Z$.
 To linear
 order we find
that
\beq
Z = -2 D_\pd H^\pp D_+ \chi + 2 D_- H^\mm D_\md \chi +
 i H^\mm \di_\mm \chi
   - i H^\pp \di_\pp \chi  ~~,
\eeq
and therefore we can express $\calX$ and, in a similar fashion,
 $\bar{\calX}$ as
\bea
\calX &=& [1-2 D_\pd H^\pp D_+ + 2 D_- H^\mm D_\md  +
 i H^\mm \di_\mm
   - i H^\pp \di_\pp] \chi  + {\cal O}(H^2) \non \\
\bar{\calX} &=& [1+2 D_+ H^\pp D_\pd - 2 D_\md H^\mm D_-  -
            i H^\mm \di_\mm  + i H^\pp \di_\pp] \chib + {\cal O}(H^2) ~~.
\eea

{\bf Acknowledgments}
The impetus for much of this work came from our interaction
with Jim Gates.
We wish to thank him for discussions, suggestions, and
general insights into many aspects of superspace and its
geometry.
We also thank Nathan Berkovits for discussions concerning
 the existence of
chiral and twisted chiral measures  in the $U_V(1) \otimes
 U_A(1)$ theory.
M.E.W. thanks the Physics Department of Queen's University
 for hospitality.
\newpage

\appendix{\Huge{\bf Appendices}}
\section{Components of covariant derivatives}
\setcounter{equation}{0}

We determine in this Appendix  the components of supergravity
 covariant
 derivatives.
  We restrict ourselves to the minimal $U_A(1)$ theory by setting
 $\Sigma_A =0$.
In defining components we follow the philosophy and methods
 described in
{\em Superspace}, section 5.6 \cite{Superspace}.
  By definition of Wess-Zumino gauge, the expressions for the
 covariant
derivatives evaluated at $\th=0$ are:
\bea
\Dela | &=& \di_\a   \non \\
\Del_a | &=& \bD_a + \psi_a^{\a} \Dela| + \psi_a^{\ad} \Delad|
 \label{Del|}
 \non\\
         &=& \bD_a + \psi_a^{\a} \di_{\a} + \psi_a^{\ad} \di_{\ad} ~~.
\eea
These expressions also serve to define the gravitino fields.

The  component connections are defined by projection:
\bea
        \F_A | &=& \vf_A   \non\\
\S_A' | &=& A' _A      \non \\
        \S_A | &=& V_A
\eea
and
\bea
      \O_A | &=& \o_A  \non \\
      \G_A | &=& \g_A ~~~.
\eea
Note that
\beq
\vf_A = \half (\o_A + \g_A) ~~{\rm and} ~~  A' _A = \sihalf
(\o_A - \g_A)  ~~.
\eeq
In the $U_A(1)$ theory we define $\bD_a$ as the fully covariant
gravitational
 derivative with a
Lorentz
connection  $\varphi_A$ that includes, in addition to the ordinary
 connection,
 extra terms
that are bilinear in the gravitini $\psi_a^{\a}$, $\psi_a^{\dot{\a}}$.
 Specifically, $\bD_a$ is defined to
be
\bea
  \bD_a &=& e_a + {\vf}_a {\cal M} +  A' _a  {\cal Y}'  \non \\
        &=& e_a + \o_a M + \g_a \Mb ~~. \label{Ddef}
\eea

 We also introduce the ordinary gravitational
covariant derivative (without $U(1)$ connection or gravitino
 bilinears),
 denoted by ${\cal D}_a$
\beq
  {\cal D}_a = e_a + {\vp}_a {\cal M}  ~~~. \label{grav}
\eeq

   We  also need expressions for
the higher $\th$ components of the covariant derivatives.
  The $\th^{\a}$ component of $\Del_{\b}$ is defined by (c.f.
{\em Superspace} sec. 5.6.b)
\beq
\Dela \Del_{\b}| = \half \{\Dela , \Del_{\b}\}|  \label{alphbeta}
\eeq
while the $\th^{\a}$ component of $\Del_b$ is
\bea
\Dela \Del_b | &=& [\Dela , \Del_b]| + \Del_b \Dela | \non \\
               &=& [\Dela , \Del_b]| + \bD_b \Dela | + \psi_b^{\b} \di_{\b}
                   \Dela | + \psi_b^{\bd} \di_{\bd} \Dela | ~~~. \label{abeta}
\eea
 From \reff{alphbeta} and (2.3)  we obtain the following results:
\bea
\Delp \Delp | &=& \half \{\Delp , \Delp \} = 0  \non \\
\Delm \Delm | &=& \half \{\Delm , \Delm \} = 0 \non\\
\Delp \Delpd | &=& \half \{\Delp , \Delpd \} = \sihalf \Delpp \non\\
\Delm \Delmd | &=& \half \{\Delm , \Delmd \} = \sihalf \Delmm\non\\
\Delp \Delm | &=& \half \{\Delp , \Delm \} = - \half \Rb | \Mb \non\\
\Delp \Delmd | &=& \half \{\Delp , \Delmd \} = 0 \nonumber\\
\Delm \Delpd | &=& \half \{\Delm , \Delpd \} = 0 ~~,
\eea
and from \reff{abeta}, we derive the series of identities that
 appears below:
\bea
\Delp \Delpp | &=&  [\Delp, \Delpp]| + \Delpp \Delp | \non \\
&=& \bDpp \Delp| + \psi_{\pp}^- \Delm \Delp| + \psi_{\pp}^{\pd}
 \Delpd \Delp|
     \non \\
&=& \bDpp \Delp| - \half \psi_{\pp}^- \Rb| \Mb +
 \sihalf \psi_{\pp}^{\pd}
   (\bDpp + \psi_{\pp}^{\a} \Del_{\a} + \psi_{\pp}^{\ad} \Del_{\ad})|
 \nonumber \\
\Delm \Delpp | &=&  [\Delm, \Delpp]| + \Delpp \Delm | \non \\
&=& \bDpp \Delm| + \psi_{\pp}^+ \Delp \Delm| + \psi_{\pp}^{\md}
 \Delmd \Delm |
    + \sihalf \Rb \Delpd | -i(\Delpd \Rb) \Mb | \non \\
&=& \bDpp \Delm| - \half \psi_{\pp}^+ \Rb| \Mb + \sihalf
 \psi_{\pp}^{\md}
   (\bDmm + \psi_{\mm}^{\a} \Del_{\a} + \psi_{\mm}^{\ad} \Del_{\ad})| +
   \sihalf \Rb \Delpd | -i(\Delpd \Rb) \Mb | \non \\
  \nonumber \\
\Delpd \Delpp |
&=& \bDpp \Delpd| + \half \psi_{\pp}^{\md} R| M + \sihalf \psi_{\pp}^+
   (\bDpp + \psi_{\pp}^{\a} \Del_{\a} + \psi_{\pp}^{\ad} \Del_{\ad})|
 \nonumber \\
\Delmd \Delpp |
&=& \bDpp \Delmd| + \half \psi_{\pp}^{\pd} R| M +
 \sihalf \psi_{\pp}^-
   (\bDmm + \psi_{\mm}^{\a} \Del_{\a} + \psi_{\mm}^{\ad}
 \Del_{\ad})| -
   \sihalf R \Delp | +i(\Delp R) M | \non \\
  \nonumber \\
\Delp \Delmm |
&=& \bDmm \Delp| - \half \psi_{\mm}^- \Rb| \Mb +
\sihalf \psi_{\mm}^{\pd}
   (\bDpp + \psi_{\pp}^{\a} \Del_{\a} + \psi_{\pp}^{\ad} \Del_{\ad})| -
   \sihalf \Rb \Delmd | -i(\Delmd \Rb) \Mb | \non \\
   \nonumber\\
\Delm \Delmm |
&=& \bDmm \Delm| - \half \psi_{\mm}^+ \Rb| \Mb + \sihalf
\psi_{\mm}^{\md}
   (\bDmm + \psi_{\mm}^{\a} \Del_{\a} + \psi_{\mm}^{\ad} \Del_{\ad})|
 \nonumber\\
\Delpd \Delmm |
&=& \bDmm \Delpd| + \half \psi_{\mm}^{\md} R| M +
\sihalf \psi_{\mm}^+
   (\bDpp + \psi_{\pp}^{\a} \Del_{\a} + \psi_{\pp}^{\ad} \Del_{\ad})| +
   \sihalf R \Delm | +i(\Delm R) M | \non \\
   \nonumber\\
\Delmd \Delmm |
&=& \bDmm \Delmd| + \half \psi_{\mm}^{\pd} R| M +
 \sihalf \psi_{\mm}^-
   (\bDmm + \psi_{\mm}^{\a} \Del_{\a} + \psi_{\mm}^{\ad}
\Del_{\ad})|
\eea
where we have used the commutation relations
\bea
{[}\Delp, \Delpp] &=& 0 ~~~, ~~~[\Delm, \Delmm ] = 0  \non\\
{[}\Delpd, \Delpp ] &=& 0 ~~~, ~~~[\Delmd, \Delmm ] = 0 \non\\
{[}\Delp, \Delmm ] &=& - \sihalf \Rb \Delmd  -i(\Delmd \Rb)
 \Mb \non \\
{[}\Delpd, \Delmm ] &=&  \sihalf R \Delm  +i(\Delm R) M \non\\
{[}\Delm, \Delpp ] &=&  \sihalf \Rb \Delpd  -i(\Delpd \Rb)
 \Mb \non\\
{[}\Delmd, \Delpp ] &=&  -\sihalf R \Delp  +i(\Delp R) M
\label{comrel}
\eea
and also
\bea
[\Delpp, \Delmm ] &=&  \half (\Delp R) \Delm +\half (\Delm R)
 \Delp
        - \half (\Delpd \Rb) \Delmd - \half (\Delmd \Rb) \Delpd \non \\
&& -  \half R \Rb \Mb - \half R \Rb M + (\Delsq R) M -(\Delbsq \Rb)
\Mb
\label{DppDmm} ~~.
\eea
We also observe that
\bea
\Delpp \Delmm | &=& (\bDpp + \psi_{\pp}^{\a} \Del_{\a} +
                               \psi_{\pp}^{\ad} \Del_{\ad}) \Delmm| \non  \\
     &=& \Delpp| \Delmm| + \psi_{\pp}^{\a} \Del_{\a} \Delmm| +
                               \psi_{\pp}^{\ad} \Del_{\ad} \Delmm| ~~,
\eea
so that
\beq
[\Delpp, \Delmm] | = [\Delpp |, \Delmm |] +  \psi_{\pp}^{\a}
\Del_{\a} \Delmm|
   + \psi_{\pp}^{\ad} \Del_{\ad} \Delmm| - \psi_{\mm}^{\a}
 \Del_{\a} \Delpp |
   - \psi_{\mm}^{\ad} \Del_{\ad} \Delpp|  ~.
\eeq
The first term can be expanded as
\bea
[\Delpp |, \Delmm |] &=& [\bDpp +  \psi_{\pp}^{\a} \di_{\a} +
 \psi_{\pp}^{\ad}
    \di_{\ad}, \bDmm + \psi_\mm^{\b} \di_{\b} + \psi_\mm^{\bd}
\di_{\bd}]
    \non \\
&=& [\bDpp, \bDmm] + \bD_{[\pp} (\psi_{\mm]}^{\a} \di_{\a}) +
      \bD_{[\pp} (\psi_{\mm]}^{\ad} \di_{\ad}) \nonumber\\
&=& [\bDpp, \bDmm] + (\bD_{[\pp} \psi_{\mm]}^{\a}) \di_{\a} +
      (\bD_{[\pp} \psi_{\mm]}^{\ad}) \di_{\ad} \non \nonumber\\
&& + \half \o_{[\pp} (\psi_{\mm]}^+ \di_+ - \psi_{\mm]}^- \di_-)
    + \half \g_{[\pp} (\psi_{\mm]}^\pd \di_\pd - \psi_{\mm]}^\md
 \di_\md)~~.
\eea
Substituting this into $[\Delpp, \Delmm] |$, we get
\bea
\lefteqn{[\Delpp, \Delmm] | = [\bDpp, \bDmm] + \bD_{[\pp}
(\psi_{\mm]}^{\a}
     \di_{\a})
    +  \bD_{[\pp} (\psi_{\mm]}^{\ad} \di_{\ad}) } \non \\
 && \hskip 6em + \psi_{\pp}^{\a} \Del_{\a} \Delmm|
    + \psi_{\pp}^{\ad} \Del_{\ad} \Delmm| - \psi_\mm^{\a} \Del_{\a}
 \Delpp |
     - \psi_\mm^{\ad} \Del_{\ad} \Delpp| \non \\
 &=& [\bDpp, \bDmm] + \bD_{[\pp} (\psi_{\mm]}^{\a} \di_{\a}) +
      \bD_{[\pp} (\psi_{\mm]}^{\ad} \di_{\ad}) \non \\
 && +  \psi_{\pp}^+ \left[\Delmm \di_+ + \sihalf \psi_\mm^{\pd}
(\bD_{\pp} +
      \psi_{\pp}^{\a} \di_{\a} + \psi_{\pp}^{\ad} \di_{\ad}) - \half
      \psi_\mm^- \Rb \Mb - \sihalf \Rb \di_{\md} -
i (\Delmd \Rb) \Mb\right] \non \\
 && + \psi_{\pp}^- \left[\bDmm \di_- - \half \psi_\mm^+
\Rb \Mb + \sihalf \psi_\mm^{\md}
   (\bDmm + \psi_\mm^{\a} \di_{\a} + \psi_\mm^{\ad} \di_{\ad})\right]
\non \\
  && + {\rm~6~other~similar~terms~}. \label{DppDmm|}
\eea
   If we compare this now with \reff{DppDmm} (which
is true in general, not just at $\th=0$), we can obtain expressions
for the
various derivatives of $R$ and $\Rb$, evaluated at $\th=0$.
 By comparing
the coefficients of $\Delp | = \di_+$ on both sides, for example,
 we get
\beq
 \half \Delm R | = \bD_{[\pp} \psi_{\mm]}^+ + \sihalf
\psi_\mm^{\md} R| +
 i \psi_{\pp}^- \psi_\mm^{\md} \psi_\mm^+ + i \psi_{\pp}^{\pd}
 \psi_\mm^+
\psi_{\pp}^+
 + i \psi_{\pp}^{\md} \psi_\mm^- \psi_\mm^+ ~~ .
\eeq
In the same fashion
\bea
 \half \Delp R | &=& \bD_{[\pp} \psi_{\mm]}^- + \sihalf
\psi_{\pp}^{\pd} R| +
 i \psi_{\pp}^+ \psi_\mm^{\pd} \psi_{\pp}^- + i \psi_{\pp}^-
 \psi_\mm^{\md}
\psi_\mm^-
 + i \psi_{\pp}^{\pd} \psi_\mm^+ \psi_{\pp}^- \nonumber \\
- \half \Delpd \Rb | &=& \bD_{[\pp} \psi_{\mm]}^{\md} - \sihalf
\psi_{\pp}^+
\Rb|
+
i \psi_{\pp}^+ \psi_\mm^{\pd} \psi_{\pp}^{\md} + i \psi_{\pp}^{\md}
 \psi_\mm^+
\psi_{\pp}^{\md}
 + i \psi_{\pp}^{\md} \psi_\mm^- \psi_{\mm}^{\md} \nonumber \\
- \half \Delmd \Rb | &=& \bD_{[\pp} \psi_{\mm]}^{\pd} - \sihalf
\psi_\mm^- \Rb|
+
 i \psi_{\pp}^+ \psi_\mm^{\pd} \psi_{\pp}^{\pd} + i \psi_{\pp}^{\md}
 \psi_\mm^-
\psi_\mm^{\pd}
 + i \psi_{\pp}^- \psi_\mm^{\md} \psi_\mm^{\pd} \nonumber\\
(-\half R \Rb + \Delsq R)| &=& [\bDpp, \bDmm]_M - i \psi_\mm^{\md}
\Delp R|
 +i \psi_{\pp}^{\pd} \Delm R| - \psi_\mm^{\md} \psi_{\pp}^{\pd} R|
 - \psi_\mm^{\pd} \psi_{\pp}^{\md} R|  \non \\
  \nonumber \\
(-\half R \Rb - \Delbsq \Rb)| &=& [\bDpp, \bDmm]_{\Mb} - i
\psi_{\pp}^+ \Delmd
\Rb|
+i \psi_\mm^- \Delpd \Rb| - \psi_{\pp}^+ \psi_\mm^- \Rb|  +
 \psi_\mm^+
\psi_{\pp}^-
\Rb|  \non  \\
\eea

\vspace{0.2cm}

 We define  the ``component" torsion   and curvature by
\beq
[\bDpp, \bDmm] = {t_{\pp \mm}}^a \bD_a + r_{\pp \mm} M +
 \bar{r}_{\pp \mm} \Mb
{}~~. \label{dd2}
\eeq
Substituting \reff{Ddef} into \reff{dd2} we have
\bea
{[}\bDpp, \bDmm] &=& [e_\pp, e_\mm] + e_{[\pp} \o_{\mm]} M +
 e_{[\pp} \g_{\mm]}
\Mb
  \non \\
&&- \half \o_{\{\mm} e_{\pp\}} - \half \g_{\{\pp} e_{\mm\}} - \o_\pp
 \o_\mm M
   \non \\
&&+ \g_\pp \g_\mm \Mb - \half \g_{\{\pp} \o_{\mm \}} M -
    \half \o_{\{\mm} \g_{\pp \}} \Mb ~~~.   \label{dd}
\eea
The anholonomy coefficients are defined as usual by
 $[e_\pp, e_\mm] =
{C_{\pp \mm}}^a e_a$. Comparing \reff{dd2} with \reff{dd},
we find the torsions
\bea
{t_{\pp \mm}}^\pp &=& {C_{\pp \mm}}^\pp - \half(\o + \g)_\mm
\label{t}  \non\\
{t_{\pp \mm}}^\mm &=& {C_{\pp \mm}}^\mm - \half(\o + \g)_\pp  ~~.
\eea
and the curvatures
\bea
r_{\pp \mm} &=& e_{[\pp} \o_{\mm]} - {C_{\pp \mm}}^\pp \o_\pp -
 {C_{\pp \mm}}
  ^\mm \o_\mm  \non\\
{\bar{r}}_{\pp \mm} &=& e_{[\pp} \g_{\mm]} - {C_{\pp \mm}}^\pp
 \g_\pp
   - {C_{\pp \mm}}^\mm \g_\mm ~~.
\eea

The full  superspace torsions are defined in the standard way,
 $[\Del_A,
 \Del_B \}
= {T_{AB}}^C \Del_C + R_{AB}M + \Rb_{AB}\Mb$.  Therefore,
from \reff{DppDmm|}
at $\th = 0$ and using \reff{Del|}, we obtain
\bea
{T_{\pp \mm}}^\pp|&=& {t_{\pp \mm}}^\pp + i (\psi_\pp^+
\psi_\mm^\pd +
       \psi_\pp^\pd \psi_\mm^+) \label{Tpp}  \non\\
{T_{\pp \mm}}^\mm |&=& {t_{\pp \mm}}^\mm + i (\psi_\pp^-
 \psi_\mm^\md +
       \psi_\pp^\md \psi_\mm^-)  \non \\
{T_{\pp \mm}}^+| &=& \bD_{[\pp} \psi_{\mm]}^+ +\sihalf
\psi_\mm^\md R|
      + i \psi_\pp^- \psi_\mm^\md \psi_\mm^+ + i \psi_\pp^\pd
\psi_\mm^+
      \psi_\pp^+ + i \psi_\pp^\md \psi_\mm^- \psi_\mm^+\non \\
{T_{\pp \mm}}^- |&=& \bD_{[\pp} \psi_{\mm]}^- +\sihalf
\psi_\pp^\pd R|
      + i \psi_\pp^+ \psi_\mm^\pd \psi_\pp^- + i \psi_\pp^-
 \psi_\mm^\md
      \psi_\mm^- + i \psi_\pp^\pd \psi_\mm^+ \psi_\pp^-  ~~.
\eea
However, from \reff{DppDmm},  ${T_{\pp \mm}}^a = 0$.
 Combining this
with \reff{t} and  \reff{Tpp}, we obtain
\bea
\vf_\pp &=& \half (\o + \g)_\pp \non \\
&=& {C_{\pp \mm}}^\mm + i (\psi_\pp^- \psi_\mm^\md +
\psi_\pp^\md
\psi_\mm^-)
\non \\
&=& \vp_\pp + i (\psi_\pp^- \psi_\mm^\md + \psi_\pp^\md
 \psi_\mm^-)
\label{conn1} \\
\vf_\mm &=& \half (\o + \g)_\mm \non \\
&=& {C_{\pp \mm}}^\pp + i (\psi_\pp^+ \psi_\mm^\pd +
\psi_\pp^\pd \psi_\mm^+)
\non \\
&=& \vp_\mm + i (\psi_\pp^+ \psi_\mm^\pd + \psi_\pp^\pd
\psi_\mm^+)
\label{conn2}
\eea
for the full component Lorentz connection, written in terms of the
ordinary component gravitational connection, ${\vp}_a$, plus
the gravitino
terms
mentioned previously.

\section{Components of Chiral Superfield Derivatives}
\setcounter{equation}{0}

We determine in this  Appendix the  $\th =0$ components of the
higher spinor derivatives of chiral superfields, as they
appear in the
component action of  subsection 4.1. The procedure is
 straightforward.
For example, we write
\bea
\Delp \Delbsq \bar{\F}| &=& [ \Delp, \Delbsq ]\bar{\F}| =
 i \Delpp \Delmd
 \bar{\F}|
\nonumber\\
&=& i(\bDpp + \psi_\pp^{\a} \Del_{\a} + \psi_\pp^{\ad}
\Del_{\ad} )\Del_\md
 \bar{\F}|
\nonumber\\
&=&i \bDpp \Delmd \bar{\F} - \psi_\pp^- \Delpp \bar{\F}
+i \psi_\pp^{\pd}
 \Del_\pd
\Del_\md \bar{\F}| \nonumber\\
&=&i \bDpp \Delmd \bar{\F} - \psi_\pp^-( \bDpp +
 \psi_\mm^\pd \Del_\pd +
\psi_\mm^\md \Del_\md ) \bar{\F}  +
i \psi_\pp^{\pd} \Del_\pd
\Del_\md \bar{\F}|  \nonumber\\
&=& i \bDpp \psi_\md - \psi_\pp^- \bDmm \bar{\f} -
\psi_\pp^-
\psi_\mm^\pd \psi_\pd - \psi_\pp^- \psi_\mm^\md \psi_\md +
 \psi_\pp^\pd
\bar{F}
\eea
where we have used the definition of components of
covariant derivatives,
as discussed in Appendix A, and the definition of components
 of $\F$.

In a similar manner  we find
\beq
\Delm \Delbsq \bar{\F}| = -i \bDmm \psi_\pd - \psi_\mm^+
 \bDpp \bar{\f} +
\psi_\mm^+
\psi_\pp^\md \psi_\md + \psi_\mm^+ \psi_\pp^\pd \psi_\pd +
 \psi_\mm^\md
\bar{F}  ~~.
\eeq
Next, using the commutation relations and the definition of the
 components of
$R$, we write
\bea
\Delsq \Delbsq \bar{\F}| &=& \Delpp \Delmm \bar{\F}| +
 \half(\Delpd \Rb)\Delmd
\bar{\F}| + \half \Rb \Delbsq \bar{\F}| \nonumber\\
&=&\Delpp \Delmm \bar{\F}| +\half \Rb \Delbsq \bar{\F}| \non \\
&&- \left[\bD_{[\pp} \psi_{\mm]}^{\pd} - \sihalf \psi_\mm^- \Rb|
+ i \psi_{\pp}^+ \psi_\mm^{\pd} \psi_{\pp}^{\pd} + i \psi_{\pp}^{\md}
\psi_\mm^-
\psi_\mm^{\pd}
 + i \psi_{\pp}^- \psi_\mm^{\md} \psi_\mm^{\pd}\right]
\Delmd \bar{\F}|~~.  \nonumber\\
\eea
Finally,
\bea
\Delpp \Delmm \bar{\F}|
&=& ( \bDpp +\psi_\pp^\a \Del_\a +\psi_\pp^\ad \Del_\ad )\Delmm
\bar{\F}|
\nonumber\\
&=& \bDpp (\bDmm + \psi_\mm^\ad \Del_\ad) \bar{\F}|
-  \ihalf \psi_\pp^+ \Rb \Delmd \bar{\F} | \non  \\
&& + \psi_\pp^\pd [ \bDmm \Delpd + i \psi_\mm^+( \bDpp +
 \psi_\pp^\ad \Delad)
- \psi_\mm^\md \Delbsq ] \bar{\F} | \non \\
&& + \psi_\pp^\md [ \bDmm \Delmd + i \psi_\mm^-( \bDmm +
 \psi_\mm^\ad \Delad)
+ \psi_\mm^\pd \Delbsq ] \bar{\F}| \non \\
&=& \bDpp \bDmm \bar{\f} + \bDpp \psi_\mm^\pd \psi_\pd +
\bDpp \psi_\mm^\md
\psi_\md
+ \psi_\pp^\pd \bDmm \psi_\pd + \psi_\pp^\md  \bDmm
\psi_\md \non \\
&&- \ihalf \psi_\pp^+ \bar{B} \psi_\md + i \psi_\pp^\pd
\psi_\mm^+ \bDpp
\bar{\f}
+i \psi_\pp^\md \psi_\mm^- \bDmm \bar{\f} \non \\
&& + i \psi_\pp^\pd \psi_\mm^+ \psi_\pp^\md \psi_\md
+ i \psi_\pp^\md \psi_\mm^- \psi_\mm^\md \psi_\md
+ i \psi_\pp^\md \psi_\mm^- \psi_\mm^\pd \psi_\pd \non \\
&& -i \psi_\pp^\md \psi_\mm^\pd \bar{F} +i \psi_\pp^\pd
\psi_\mm^\md \bar{F} ~~.
\eea

We now have all the ingredients for computing the component
action for
the chiral multiplet, and in particular the terms involving the
auxiliary
fields which are needed for the computation in subsection 4.1 .
 Note that  the
component  covariant derivative appearing above still
contains the
connection involving gravitino bilinears and a $U(1)$
 gauge field.

\section{Derivation of Twisted Chiral Projection Formula}
\setcounter{equation}{0}

In subsection 4.2 we have  presented the projection formula
 for the
component action in terms of a twisted chiral projector. The
manipulations that
lead to \reff{twistdensity} are rather baroque, and we can
 only try to gently
 guide
the enterprising reader (or, more likely, the hapless graduate
 student) through
some of the steps.

We consider the right hand side of \reff{genident} and study
separately the
 various
terms  after applying the identity to an arbitrary superspace
lagrangian,
integrating with $\int d^2x e^{-1}$ and evaluating at $\th =0$.

We begin by considering the term
\bea
&& \frac{1}{4} (\Delm R) \Delp { \cal L}| \nonumber \\
&=&
\frac{1}{2} \left[ \bD_{[ \pp} \psi_{\mm ]}^+ +\ihalf
\psi_\mm^\md B
+i \psi_\pp^- \psi_\mm^\md \psi_\mm ^+ + i \psi_\pp^\pd
 \psi_\mm^+ \psi_\pp^+
+i \psi_\pp^\md \psi_\mm^- \psi_\mm^+ \right] \Delp {\cal L}|
\eea
where we have substituted the $\th =0$ expression for $\Delm R$.
Now,
\bea
\bDpp \psi_\mm^+ &=& (e_\pp + \omega_\pp M + \g_\pp \bar{M})
\psi_\mm^+
\nonumber\\
&=& ( \calDpp +\ihalf A_\pp ' )\psi_\mm^+ -\frac{3i}{2}(\psi_\pp^-
 \psi_\mm^\md
+ \psi_\pp^\md \psi_\mm^-)\psi_\mm^+
\eea
and
\bea
\bDmm \psi_\pp^+ &=& (e_\mm + \omega_\mm M + \g_\pp \bar{M})
\psi_\pp^+
\nonumber\\
&=& ( \calDmm +\ihalf A_\mm ' )\psi_\pp^+ +\frac{i}{2}\psi_\pp^\pd
\psi_\mm^+
 \psi_\pp^+
\eea
where we have used the expressions for the connections in
 \reff{conn1},
\reff{conn2}
and the definition of the gravitational covariant derivative in
 \reff{grav}.
The term above becomes
\bea
\frac{1}{4} (\Delm R) \Delp {\cal L} |&=&
\frac{1}{2} \left[  \calD_{ [\pp} \psi_{\mm ]}^+ +\frac{i}{2}
A_{[ \pp}'  \psi_{ \mm ] }^+
  +\frac{i}{2} \psi_\mm^\md B \right.  \nonumber\\
&& \left. - \ihalf  \psi_\pp^- \psi_\mm^\md  \psi_\mm^+ + \ihalf
 \psi_\pp^\pd  \psi_\mm^+
\psi_\pp^+
-\ihalf \psi_\pp^\md \psi_\mm^- \psi_\mm^+ \right] \Delp {\cal L}|
\eea
Furthermore, since we are inside a $d^2x e^{-1}$ integral, the
 ordinary
 gravitational
derivative $\calD$ can be integrated by parts.

The other terms of a similar type are manipulated in the same
fashion.

We consider now
\beq
\frac{1}{2} \Delpp \Delmm {\cal L}| = \frac{1}{2}( \bDpp +
\psi_\pp^\a \Del_\a
+\psi_\pp^\ad \Del_\ad ) \Delmm {\cal L} |
\eeq
and write out
\beq
\bDpp \Delmm {\cal L}| = [ \calDpp  -i (\psi_\pp^-\psi_\mm^\md
+\psi_\pp^\md \psi_\mm^- )
\Delmm ] {\cal L}|
\eeq
(note that $\Delmm {\cal L}$ is $U_A(1)$ neutral).   Inside the $d^2 x
 e^{-1}$ integral, the
term ${\cal D}_\pp \Delmm {\cal L}$ is a total derivative and can
 be dropped.
For the remaining terms in
the above expression we commute the $\Delmm$ operator to the
 left of the
 $\Del_\a$
operators, using the commutation relations in \reff{comrel}, so that
\bea
&&(\psi_\pp^\a \Del_\a
+\psi_\pp^\ad \Del_\ad ) \Delmm {\cal L} |   \nonumber\\
&&=( \psi_\pp^\a \Delmm \Del_\a + \psi_\pp^\ad \Delmm \Del_\ad
 -\ihalf
 \psi_\pp^+
\Rb \Del_\md +\ihalf \psi_\pp^\pd R \Del_- ) {\cal L}|
\eea
We treat the $\Delmm \Delpp {\cal L}|$ term in the same fashion.

Assembling everything, the right-hand-side of the original identity
becomes
\bea
&& \half (-R \Del^2 + \Rb \Delb^2) {\cal L}| \nonumber\\
&&+ \half \left[ \psi_\mm^+ (\Delpp -\calDpp +\ihalf A_\pp ')
-\psi_\pp^+(\Delmm - \calDmm +\ihalf A_\mm ') +2\psi_\pp^+
 \Delmm
\right] \Delp
{\cal L}| \nonumber\\
&&+\half \left[ -\ihalf \psi_\pp^-\psi_\mm^\md \psi_\mm^+ +
\ihalf \psi_\pp^\pd
\psi_\mm^+ \psi_\pp^+
-\ihalf \psi_\pp^\md \psi_\mm^- \psi_\mm^+ \right] \Delp {\cal L}|
\nonumber\\
&&+( {\rm similar ~ terms~ with} ~~ \Delm {\cal L} ~~{\rm etc.})
\nonumber\\
&&+\left[\ihalf (\psi_\pp^+ \psi_\mm^\pd +\psi_\pp^\pd
\psi_\mm^+ )\Delpp
-\ihalf  (\psi_\pp^- \psi_\mm^\md +\psi_\pp^\md \psi_\mm^-)
\Delmm \right]
 {\cal L}|
\eea
All the terms of the form $B\Del_\a {\cal L}$  have cancelled.

Next we work out
\bea
\half \Delpp \Delp {\cal L}| &=& \half (\bDpp
+\psi_\pp^\a \Del_\a +\psi_\pp^\ad \Del_\ad  )\Delp
{\cal L}|  \\
&=&\half \left[ \calDpp -\ihalf A_\pp '  +\ihalf
(\psi_\pp^- \psi_\mm^\md
+\psi_\pp^\md \psi_\mm ^- ) +\psi_\pp^\a \Del_\a +
\psi_\pp^\ad \Del_\ad \right]
\Delp {\cal L}| \non
\eea
and, in a similar manner,  $\half \Delmm \Delp {\cal L}|$.

Substituting into (C.8), we find that all three-gravitino
terms cancel,
and the two-gravitino terms in the last line are also cancelled.

We obtain, finally, for the right-hand-side of the identity (4.11)
\bea
&&
\left\{ [\psi_\pp^+\psi_\mm^- +\psi_\pp^-\psi_\mm^+ -
\half R]\Del^2
+ [\psi_\pp^\pd \psi_\mm^\md +\psi_\pp^\md \psi_\mm^\pd  +
 \half\Rb ]\Del^2
 \right.\nonumber\\
&&\left. + (\psi_\pp^+ \psi_\mm^\md +\psi_\pp^\md
\psi_\mm^+) \Delp \Delmd
+(\psi_\pp^\pd \psi_\mm^-+\psi_\pp^- \psi_\mm^\pd)
 \Delpd \Delm   \right.
\nonumber\\
&&\left.  +\psi_\pp^+ \Delmm \Delp +\psi_\mm^- \Delpp
\Delm
+\psi_\pp^\pd \Delmm \Delpd +\psi_\mm^\md \Delpp
\Delmd \right\} {\cal L}|
\eea
which can be rearranged into the form given in the main text.

\end{document}